\documentclass[%
reprint,
superscriptaddress,
 amsmath,amssymb,
 prl,
]{revtex4-2}

\usepackage[utf8]{inputenc}
\usepackage{amsmath} 
\usepackage[colorlinks,linkcolor=blue,citecolor=magenta]{hyperref}
\usepackage{float}
\usepackage{bbding}
\usepackage[version=3]{mhchem} 
\usepackage{siunitx} 
\usepackage{graphicx}
\usepackage{color}

\newcommand{\red}[1]{ {\color{black}{#1}} }

\newcommand{\commentCWS}[1]%
{\textsf{\textcolor{blue}{#1$^{\mathrm{CWS}}$}}}

\begin{document}

\title{Exploring glassy dynamics with Markov state models from graph dynamical neural networks}
\author{Siavash Soltani}
\affiliation{Department of Materials Engineering, The University of British Columbia, Vancouver, BC, Canada V6T 1Z4}
\author{Chad W.~Sinclair}
\affiliation{Department of Materials Engineering, The University of British Columbia, Vancouver, BC, Canada V6T 1Z4}
\author{J\"org Rottler}
\affiliation{Department of Physics and Astronomy, The University of British Columbia, Vancouver, BC, Canada V6T 1Z1}
\affiliation{Stewart Blusson Quantum Matter Institute, The University of British Columbia, Vancouver, BC, Canada V6T 1Z4}

\begin{abstract}

Using machine learning techniques, we introduce a Markov state model (MSM) for a model glass former that reveals structural heterogeneities and their slow dynamics by coarse-graining the molecular dynamics into a low-dimensional feature space. The transition timescale between states is larger than the conventional structural relaxation time $\tau_\alpha$, but can be obtained from trajectories much shorter than $\tau_\alpha$. The learned map of states assigned to the particles corresponds to local excess Voronoi volume. These results resonate with classic free volume theories of the glass transition, singling out local packing fluctuations as one of the dominant slowly relaxing features. 
\end{abstract}

\maketitle

Establishing robust links between structure, dynamics and thermodynamics in amorphous materials at the nanoscale is notoriously difficult due to the complex local environment of constituent particles \cite{berthier2011theoretical,widmer2008irreversible,royall2008direct,candelier2010spatiotemporal,tanaka2019revealing}. Machine learning (ML) techniques have received abiding attention due to their powerful capabilities in dimensionality reduction and extracting dominant structural features linked causally to local mobility \cite{svm-joerg,schoenholz2016structural,cubuk2016structural,schoenholz2017relationship,schoenholz2018combining,wang2020predicting,bapst2020unveiling,yang2021machine}. Many of these methods proposed so far require supervised training on particles that were previously identified as mobile, a task that becomes extremely cumbersome in computer simulations deep in the glassy state where relaxation times are very long. Others classify particles based on structural information alone, but do not yield the relaxation times associated with the learned features \cite{boattini2020autonomously,paret2020assessing}. What is needed is a method that treats structure and dynamics on the same footing by analysing directly the molecular trajectories, `automatically' identifying the rate limiting processes and then associating particles to the corresponding spatial features. In this way, one can hope to simultaneously identify the relevant hydrodynamic variables, their relaxation timescales, and their equilibrium thermodynamics.

Exactly such a program is offered by Markov state models (MSMs)  \cite{van1992stochastic}. Importantly, MSMs enable one to extract dynamical information from an ensemble of trajectories each much shorter than the timescale of interest. These models have a long tradition in molecular biophysics \cite{swope2004describing,prinz2011markov,husic2018markov}, where they are used to capture slow conformational changes of proteins and biomolecules that occur far beyond the nanonsecond timescales of direct MD simulations \cite{voelz2010molecular}. Major progress occurred with the introduction of the variational principle for Markov processes (VAMP) \cite{noe2013variational}, which when combined with deep learning methods permits the automated construction of MSMs from molecular trajectories \cite{mardt2018vampnets}. The VAMP formulation of MSMs has been combined recently with graph dynamical networks (GDyNets) in order to infer the dynamics of atomic scale processes in various condensed matter systems \cite{xie2019graph}. In this method, the local environment of particles is mapped into a state space where the dynamics are linear, and the time scale of the limiting processes are extracted using MSMs directly trained on the dynamics of particles. 

Here, we use VAMP/GDyNets based MSMs to study the long time dynamics of a model glass forming liquid over a wide range of temperatures ranging from the liquid state down to deep in the glassy regime, where standard analysis of dynamics is extremely challenging. We construct a minimal two-state MSM that reveals relaxation processes on diffusive timescales, much longer than the main structural $\alpha$-relaxation time that is conventionally used to characterize glassy dynamics. The conversion timescale $\tau_{MSM}$ between the two states has a super-Arrhenius temperature dependence above the glass transition, but crosses over into simple Arrhenius behavior at low temperatures. The transitions between the states identified by our MSM describe local excess volume fluctuations. As assumed in classical free volume theories of the glass transition, the free energy difference between these states is purely entropic. The MSM viewpoint of glassy dynamics proposed here thus identifies local packing fluctuations as the slowest relaxing structural observable.

\textit{Two-state Markov model.}
In Markov models, the (discrete) state at time $t+\delta t$ only depends on the state at time $t$. The time evolution of the system is then captured by the transition matrix, $\boldsymbol{K}(\delta t)$ that propagates the probabilities of finding the system in a given state. Elements of the transition matrix $ K_{ij}(\delta t)$ are the probabilities of observing the system in state $j$ at $t=t+\delta t$ given that the system was in state $i$ at time $t$. The optimal feature map functions that assign this probability to each particle are learned from molecular dynamics trajectories with graph convolutional networks as described in refs. \cite{xie2019graph,mardt2018vampnets} and in Supplemental Material \cite{supp-mat}.

A two-state Markov model is completely defined by the conversion rates $k_{0 \rightarrow 1}$ and $k_{1 \rightarrow 0}$ between states 0 and 1. The nontrivial eigenvalue $\lambda_2$ of the transition matrix $\boldsymbol{K}(\delta t)$ determines the total conversion timescale,
\begin{equation}
    \tau_{MSM} = -\frac{\delta t}{\ln{|\lambda_2|}}= \frac{1}{k_{0 \rightarrow 1}+k_{1 \rightarrow 0}}.
    \label{taumsm-eq}
\end{equation}
If the process is Markovian, $\tau_{MSM}$ does not depend on the choice of $\delta t$. Therefore, to test the convergence to Markovianity, one should approximate $\tau_{MSM}$ as a function of lag time $\delta t$, see Supplemental Material Fig.~1 \cite{supp-mat}. The equilibrium distribution (eigenvector associated with $\lambda_1=1$) is given by

\begin{equation}
    v_1 =\left( \frac{k_{1 \rightarrow 0}}{k_{0 \rightarrow 1}+k_{1 \rightarrow 0}},\frac{k_{0 \rightarrow 1}}{k_{0 \rightarrow 1}+k_{1 \rightarrow 0}}\right). 
    \label{eigvec-eq}
\end{equation}

\begin{figure}[t]
    \includegraphics[width=\columnwidth]{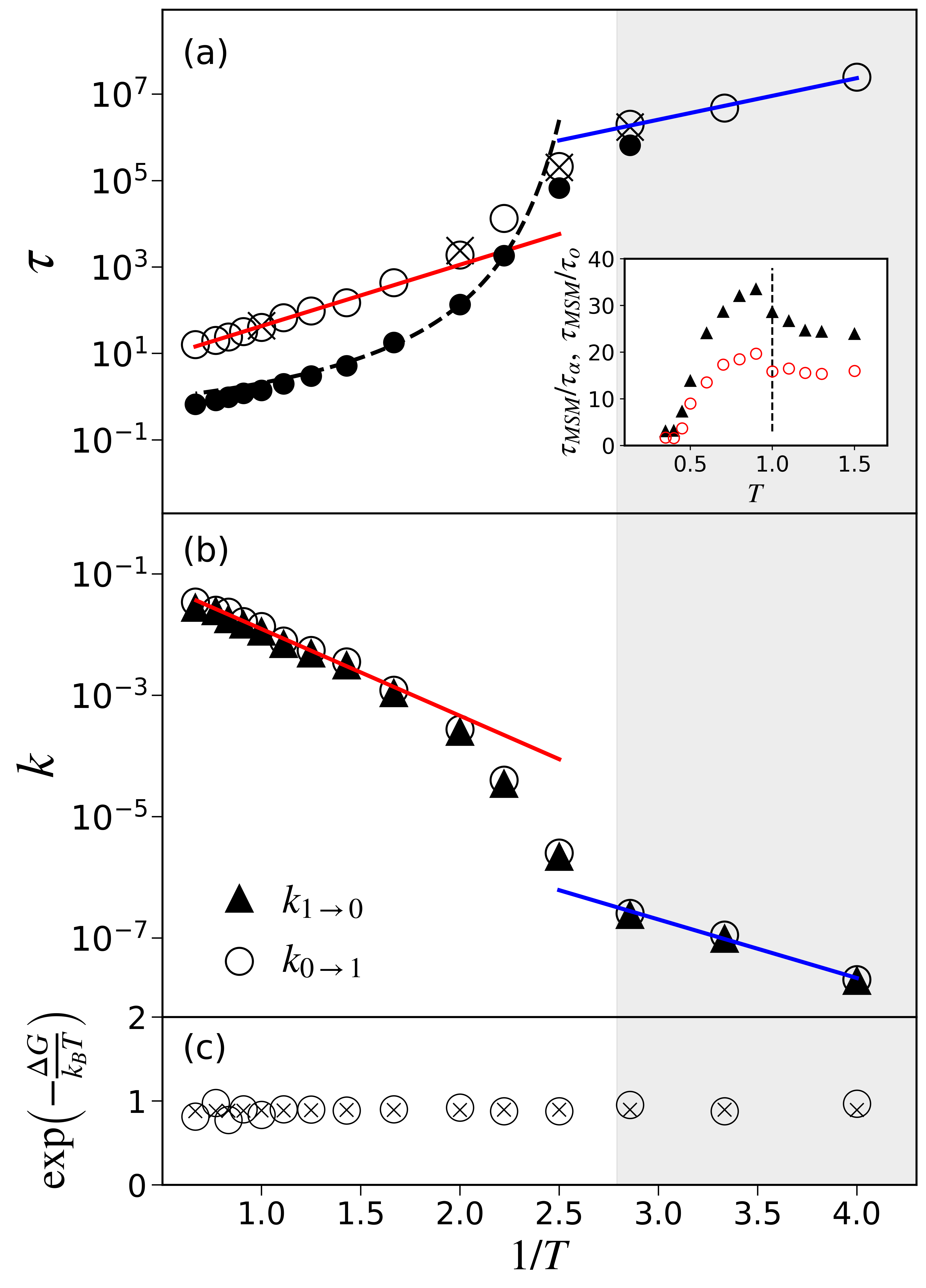} 
    \caption{(a) Markov state model relaxation times $\tau_{MSM}$ (empty circles) and $\alpha$-relaxation times $\tau_\alpha$ (filled circles) as a function of inverse temperature. \red{Crosses show the time scales obtained directly from the decay of autocorrelation functions of smoothed Voronoi volume (see below).} Dashed curve shows a Vogel–Fulcher–Tammann fit, $\tau_{\alpha} \sim \exp[A/(T-T_0)]$ for $T\ge 0.45$. The obtained $T_0$ value is 0.34. Arrhenius fits at high and low temperatures are also shown with red and blue solid lines, resulting in energy barriers of $3.27$ and \red{$2.21$} LJ energy units, respectively. In the grey shaded area, the system is out of equilibrium.
    The inset shows the ratio of $\tau_{MSM}$ to $\tau_\alpha$ (triangles) and also to the collective overlap time $\tau_O$ (circles, see text).    (b) Forward and backward reaction rates, $k_{0 \rightarrow 1}$ and $k_{1 \rightarrow 0}$ with Arrhenius fits at high and low T. (c) Free energy difference between states (circles) and ratio of mean state probabilities (crosses, see text).}
    \label{fig:msm}%
\end{figure}

\begin{figure*}[t]
  \centering
  \includegraphics[width=\textwidth]{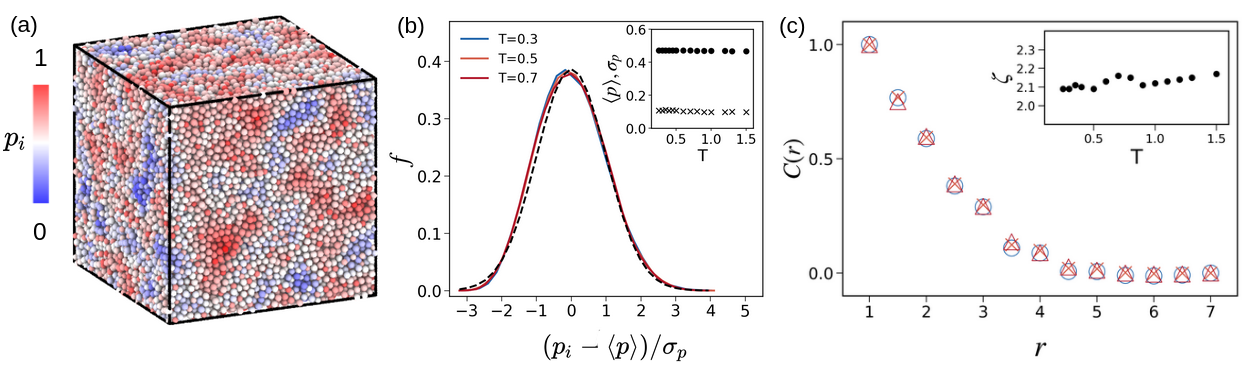}
  \caption{(a) Example of a state map at T=0.3. Particles $i$ are colored according to their probability $p_i$ of being in state 0, as assigned by the graph dynamical network. (b) Probability distribution of the normalized probability $\delta p_i = (p_i-\left<p\right>)/\sqrt{\left<( p-\left<p\right>)^2\right>}$ of majority type particles of being in state 0 at 3 different temperatures. Distributions are averaged over 100 equally spaced snapshots of the trajectory. \red{Dashed curve shows Gaussian fit. Inset shows mean (circles) and standard deviations (crosses) as a function of temperature.} (c) Normalized spatial correlation functions $C(r) = \left<\delta p(0)\delta p(r) \right>/\left<\delta p(0)\delta p(r_{min}) \right>$ of the state maps at same temperatures as in panel (b), where the average is over all particles separated by a distance $r$ and $r_{min}=1$. Measurements are obtained using \red{20} different quench realizations to desired temperatures and an iso-thermal equilibration of 500 LJ time. The snapshot at the end of this equilibration period is then used to extract the data points in (c). Inset shows the correlation length of the state map $\zeta ={\sum_{r_{min}}^{r_{cut}} rC(r)\delta r}/{\sum_{r_{min}}^{r_{cut}}{C(r)}\delta r }$ with $\delta r=0.5$ and $r_{cut}=7$. Visualizations are performed using Ovito \cite{ovito}.}
  \label{fig:states}
\end{figure*}

We focus on simulating an extremely well characterized model glass former, namely N=40,000 particles of a 80-20 binary Lennard-Jones mixture in a periodic simulation box at various temperatures ranging from $T=0.25$ to $T=1.5$ (all quantities are given in reduced units, see Supplemental Material for details \cite{supp-mat}). Feature map functions are trained on the trajectories of the majority particles at temperature $T=0.4$ and we then extract the relaxation time of the Markov process $\tau_{MSM}$ across all temperatures.  In all cases, the lag time $\delta t$ is large enough for $\tau_{MSM}$ to be independent of $\delta t$, and $\tau_{MSM}$ does not depend on the temperature at which the model is trained (see Supplemental Material \cite{supp-mat}). Results are shown in Fig.~\ref{fig:msm}(a) in an Arrhenius representation. The conventional $\alpha$-relaxation times $\tau_\alpha$  extracted from the decay of the self-intermediate scattering function $F_s({\bf q},t)=1/N\sum_i^N\cos[{\bf q}({\bf r}_i(t)-{\bf r}_i(0)]$ to $1/e$ for $|{\bf q}|=7.2$ are also shown for temperatures in the fluid and supercooled regime $T\geq 0.4$, where the system is in equilibrium, The MSM time is always larger than the $\alpha$-relaxation time by a factor 3-30 (see inset black triangles). This suggests that the process identified by the Markov model as rate limiting is not the $\alpha$-relaxation itself, but a transition which takes place over longer time scales than cage-breaking rearrangements. The MSM process is also slower than other collective processes that are typically investigated in glasses, for instance via the decay of the collective overlap function $F_O(t,a)=1/N\sum_{i,j}\Theta(a-|\vec{r}_i(t)-\vec{r}_j(0)|)$ (see inset red circles; here $a=0.3$).

The MSM model can be constructed for temperatures deep in the glassy regime $T< 0.4$ as well, where it reveals a crossover to purely Arrhenius behavior near the glass transition $(T_c\approx 0.435/T_g\approx 0.4)$. At these temperatures (grey shaded area), the glass has fallen out of equilibrium and all timescales become history (i.e.~age) dependent. It is important to realize, however, that the system does not age appreciably in the time interval needed to construct the MSM. Moreover, at the lowest temperature studied here, this time interval is only $5$\% of $\tau_{MSM}$, while direct calculation of $\tau_\alpha$ requires a trajectory of order $\tau_\alpha$ itself.

Writing the rates in the form $k_{i\rightarrow j}=\tau_0^{-1}\exp[-(G^*-G_i)/k_BT]$, where $G_i$ and $G^*$ denote the free energies in state $i$ and in the transition state, we can interpret the two-state model in an energy landscape picture. According to equations~(\ref{taumsm-eq}) and (\ref{eigvec-eq}), the rates themselves can be obtained by dividing the components of the eigenvector $v_1$ by $\tau_{MSM}$. In Fig.~\ref{fig:msm}(b) we show the `forward' and `backward' reaction rates $k_{0 \rightarrow 1}$ and $k_{1 \rightarrow 0}$, respectively. The rates almost trace each other, which suggests a very small free energy difference between the states. An Arrhenius fit to the high temperature regime reveals an activation barrier of $3.29 \pm 0.11 $ LJ energy units, together with a pre-exponential factor $\tau_0\approx 1$. In the fluid, we expect indeed an attempt frequency of order an atomic vibration (Einstein) frequency. The barrier then rises strongly as the temperature drops, signalling the approaching glass transition. In the nonequilibrium glassy state, the barrier becomes temperature independent again and settles on a somewhat smaller value of \red{$2.19 \pm 0.24$} LJ enery units, but the preexponential time is now about a factor 1000 larger. This behavior signals a large entropic penalty for rearrangement, in agreement with the notion that cooperative motion of many particles is now needed for relaxation to occur.

A final insight that can be obtained from the transition matrix concerns the free energy difference $\Delta G=G_2-G_1$ between the states, which can be inferred from the ratio of the eigenvector components (assuming the same prefactor), $k_{0 \rightarrow 1}/{k_{1 \rightarrow 0}} \approx e^{-\Delta G/k_BT}.$
Fig.~\ref{fig:msm}(c) shows that the ratio is temperature independent, from which we infer $\Delta G=0.1 k_BT$. The linear temperature dependence implies that $\Delta G$ is purely entropic in origin.

\textit{Heterogeneous state maps.}
Fig.~\ref{fig:states}(a) shows the output of the graph dynamical network for snapshots of the simulation in the glassy regime. Since the number of states is defined to be two, the network assigns a probability $p_i$ to each particle $i$ which denotes how similar the local environment of a particle is to state 0. Distinct clusters of particles in the same state emerge, reflecting a form of heterogeneity. For this projection, a separate network was trained on the dynamics of the minority particles. The assigned states merge seamlessly into those for the majority particles, while $\tau_{MSM}$ is identical (training on all particles simultaneously in the binary mixture results in a trivial partitioning by chemical identity). The timescale, $\tau_{MSM}$, can also be thought of as the time required for the distribution of $p_i$ to spatially decorrelate.

The histograms of the (normalized) state probabilities of the majority particles in Fig.~\ref{fig:states}(b) reveal broadly peaked distributions that are insensitive to temperature (inset). The average probability $\langle p_i \rangle=0.47$ agrees well with the value of the (also temperature insensitive) eigenvector component, which corresponds to the equilibrium distribution of the two-state model. These results suggest that the distributions in Fig.~\ref{fig:states}(b) represent equilibrium configurations, which they can presumably reach at high temperature where $\tau_{MSM}$ is small. The heterogeneous state maps from our MSM bears superficial similarity with the maps of a scalar structural order parameter $P$ obtained by Boattini et al.~\cite{boattini2020autonomously}, who encoded the local atomic environment into bond order parameters and classified them using a neural network auto-encoder. However, the mean of this order parameter $\langle P\rangle$ decreases with increasing temperature, while $\langle p_i \rangle$ does not. 

\begin{figure}[t]
  \centering
  \includegraphics[width=\columnwidth]{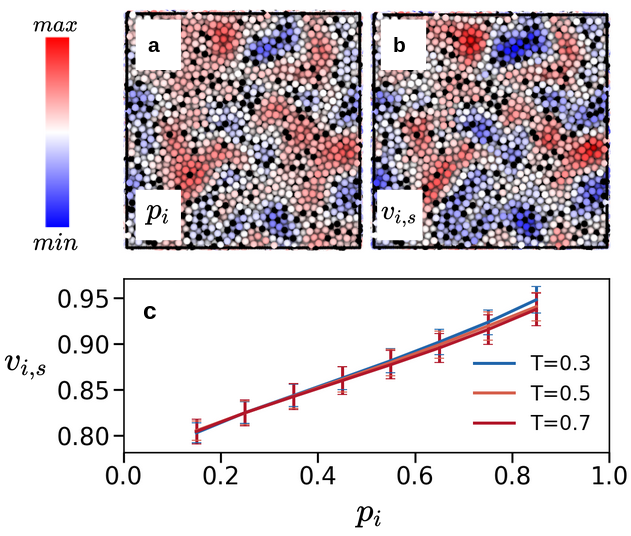}
  \caption{Comparison between (a) state of majority particles and (b) coarse grained Voronoi volume for a snapshot at $T=0.4$. Voronoi volume of particles are lightly smoothed by averaging the Voronoi volume of a particle with its neighbors within $r=\zeta$. Particles are colored using $min(p_i)=0$ and $max(p_i)=1$ in (a) and according to $min(v_{i,s})=0.81$ and $max(v_{i,s})=0.95$ in (b). Minority particles are shown in black. (c) Correlation between probability of $p_i$ and ${v}_{i,s}$ at three different temperatures. The Pearson correlation factor between $p_i$ and $v_{i,s}$ is equal to $0.88$, $0.85$ and $0.81$ at $T=0.3$, $T=0.5$ and $T=0.7$ respectively. Error bars show the standard error of the mean.}
  \label{fig:voro}
\end{figure}

In order to measure the size of the features, we compute spatial correlation functions of the state map in Fig.~\ref{fig:states}(c) and extract a characteristic correlation length. The inset of panel (c) shows the correlation length of the state map, which \red{takes an essentially temperature independent value $\zeta \approx 2.1$.} The spatial features of the Markov state map are therefore larger than those of a softness map  \cite{schoenholz2016structural}, for which a smaller correlation length of order one particle diameter has been reported \cite{cubuk2016structural}. Accordingly, one expects the lifetimes of the softness map to be shorter. \red{While dynamical correlation lengths have been reported to vary between $1-5$ in similar settings \cite{flenner2015fundamental}, the present correlation length is static, for which temperature dependence is in general weaker than for dynamical correlation lengths.}

The question is now what dynamic process the VAMP/GDyNet identifies as the slow (i.e.~hydrodynamic) mode of relaxation of the system? We find that there is a strong correlation between the state map (composed of majority particles) and local excess Voronoi volume of the same particles. To quantify this correlation, we use the standard Pearson correlation factor $\rho$, which ranges between $-1$ and $1$, with $\rho=0$ indicating no correlations.
The Pearson correlation between state maps and maps of local Voronoi volumes is near 0.5 for all temperatures. In order to reduce the inherent noise in the site-to-site variation of Voronoi volumes, we also consider a lightly coarse-grained map of Voronoi volumes, $v_{i,s}$, by averaging over particles within a range $r=\zeta$, where $\zeta$ is the correlation length in the state map, see Fig.~\ref{fig:states}(c). This local smoothing significantly increases the Pearson factor to above 0.8. Such high values imply nearly identical configurations, as can be confirmed by visual inspection of two example maps in Fig.~\ref{fig:voro}(a) and (b). Fig.~\ref{fig:voro}(c) further illustrates the link between Markov states and local Voronoi volume in the form of a nearly linear relationship between the two variables. \red{A direct measurement of the decorrelation timescale of the smoothed Voronoi map for selected temperatures agrees well with $\tau_{MSM}$ (crosses in Fig.~\ref{fig:msm}(a)).}

The equivalence of the distributions of the state probabilities and Voronoi volumes implies that the entropy difference between the two states can also be interpreted as an entropy difference of Voronoi cell volumes. This finding resonates with classic free volume theories of the glass transition. For instance, Cohen and Grest \cite{cohen1979liquid} define a local free volume as $v_f=v-v_c$, where $v_c$ is a critical volume that separates cells into liquid like and solid like cells, and the entropy of the probability distribution is the only contribution to the free energy change involved in redistributing free volume. In close analogy, our MSM describes transitions between regions of high and low excess free volume, and yields a purely entropic free energy difference between these regions. Indeed, assuming that the free volume is proportional to $p_i$, one can estimate the configurational entropy difference between the two states as $\exp{[-\Delta S/k_B]} = \langle p_i\rangle/(1-\langle p_i \rangle)$, which is in perfect agreement with the free energy estimate from the MSM, see Fig.~\ref{fig:msm}(c).

We have introduced a two-state Markov model for the study of slow dynamics in glassy systems using unsupervised machine learning techniques \red{(Results for a thre-state model are briefly discussed in Supplementary Secton 1.5)}. Following earlier work \cite{mardt2018vampnets,xie2019graph}, feature map functions were learned with graph dynamical networks (GDyNets) in combination with the variational approach for Markov processes (VAMP). The graph neural network proves to be a powerful and effective method to map the complicated local environment of particles to a state space with low dimensions, resulting in a coarse-grained description of dynamics. As a key insight of this work, the structurally agnostic GDyNets selects transitions between regions of high and low Voronoi volume as a slow mode of relaxation that can be considered Markovian. The free energy difference between the two states was found to be proportional to temperature, implying that it results from  entropy differences between the two states. Since the dynamics are linear in the state space, the relaxation towards equilibrium is purely exponential, enabling us to extract the timescales from short trajectories. As a result, the analysis can be pushed deep into the glassy regime where correlation functions probing longer length scales decay too slowly to extract timescales. Here we were able to observe very clearly the crossover from super-Arrhenius to simple Arrhenius dynamics in the energy landscape regime dominated by activated hopping below the glass transition \cite{lad2012signatures}.

A prediction of this work is that local (excess) volume fluctuations are the longest living features (i.e. the most important hydrodynamic variable) in the glass, and all other structural heterogeneities should equilibrate faster. Importantly, the structural heterogeneities are measurable at all temperatures, whereas dynamical heterogeneities disappear in the fluid regime.
Our approach is philosophically distinct from recent unsupervised learning approaches \cite{boattini2020autonomously,paret2020assessing} that use structural information alone to establish links between local order and dynamics in the supercooled regime. By including the observed dynamics, our approach instead constructs a kinetic model applicable to all temperatures that makes predictions about the thermodynamics of the underlying two-state system. \red{The additional slow timescale that emerges from our study should provide a useful benchmark for theories of the glass transition, which traditionally have focused mostly on the (cage breaking) $\alpha-$process.}

We thank Ludovic Berthier, Liesbeth Janssen, and C\'eline Ruscher for insightful comments on this work. This work was supported by the Natural Sciences and Engineering Research Council of Canada (NSERC). Computing resources were provided by ComputeCanada.



\begin{thebibliography}{28}%
\makeatletter
\providecommand \@ifxundefined [1]{%
 \@ifx{#1\undefined}
}%
\providecommand \@ifnum [1]{%
 \ifnum #1\expandafter \@firstoftwo
 \else \expandafter \@secondoftwo
 \fi
}%
\providecommand \@ifx [1]{%
 \ifx #1\expandafter \@firstoftwo
 \else \expandafter \@secondoftwo
 \fi
}%
\providecommand \natexlab [1]{#1}%
\providecommand \enquote  [1]{``#1''}%
\providecommand \bibnamefont  [1]{#1}%
\providecommand \bibfnamefont [1]{#1}%
\providecommand \citenamefont [1]{#1}%
\providecommand \href@noop [0]{\@secondoftwo}%
\providecommand \href [0]{\begingroup \@sanitize@url \@href}%
\providecommand \@href[1]{\@@startlink{#1}\@@href}%
\providecommand \@@href[1]{\endgroup#1\@@endlink}%
\providecommand \@sanitize@url [0]{\catcode `\\12\catcode `\$12\catcode
  `\&12\catcode `\#12\catcode `\^12\catcode `\_12\catcode `\%12\relax}%
\providecommand \@@startlink[1]{}%
\providecommand \@@endlink[0]{}%
\providecommand \url  [0]{\begingroup\@sanitize@url \@url }%
\providecommand \@url [1]{\endgroup\@href {#1}{\urlprefix }}%
\providecommand \urlprefix  [0]{URL }%
\providecommand \Eprint [0]{\href }%
\providecommand \doibase [0]{https://doi.org/}%
\providecommand \selectlanguage [0]{\@gobble}%
\providecommand \bibinfo  [0]{\@secondoftwo}%
\providecommand \bibfield  [0]{\@secondoftwo}%
\providecommand \translation [1]{[#1]}%
\providecommand \BibitemOpen [0]{}%
\providecommand \bibitemStop [0]{}%
\providecommand \bibitemNoStop [0]{.\EOS\space}%
\providecommand \EOS [0]{\spacefactor3000\relax}%
\providecommand \BibitemShut  [1]{\csname bibitem#1\endcsname}%
\let\auto@bib@innerbib\@empty
\bibitem [{\citenamefont {Berthier}\ and\ \citenamefont
  {Biroli}(2011)}]{berthier2011theoretical}%
  \BibitemOpen
  \bibfield  {author} {\bibinfo {author} {\bibfnamefont {L.}~\bibnamefont
  {Berthier}}\ and\ \bibinfo {author} {\bibfnamefont {G.}~\bibnamefont
  {Biroli}},\ }\bibfield  {title} {\bibinfo {title} {Theoretical perspective on
  the glass transition and amorphous materials},\ }\href@noop {} {\bibfield
  {journal} {\bibinfo  {journal} {Reviews of modern physics}\ }\textbf
  {\bibinfo {volume} {83}},\ \bibinfo {pages} {587} (\bibinfo {year}
  {2011})}\BibitemShut {NoStop}%
\bibitem [{\citenamefont {Widmer-Cooper}\ \emph {et~al.}(2008)\citenamefont
  {Widmer-Cooper}, \citenamefont {Perry}, \citenamefont {Harrowell},\ and\
  \citenamefont {Reichman}}]{widmer2008irreversible}%
  \BibitemOpen
  \bibfield  {author} {\bibinfo {author} {\bibfnamefont {A.}~\bibnamefont
  {Widmer-Cooper}}, \bibinfo {author} {\bibfnamefont {H.}~\bibnamefont
  {Perry}}, \bibinfo {author} {\bibfnamefont {P.}~\bibnamefont {Harrowell}},\
  and\ \bibinfo {author} {\bibfnamefont {D.~R.}\ \bibnamefont {Reichman}},\
  }\bibfield  {title} {\bibinfo {title} {Irreversible reorganization in a
  supercooled liquid originates from localized soft modes},\ }\href@noop {}
  {\bibfield  {journal} {\bibinfo  {journal} {Nature Physics}\ }\textbf
  {\bibinfo {volume} {4}},\ \bibinfo {pages} {711} (\bibinfo {year}
  {2008})}\BibitemShut {NoStop}%
\bibitem [{\citenamefont {Royall}\ \emph {et~al.}(2008)\citenamefont {Royall},
  \citenamefont {Williams}, \citenamefont {Ohtsuka},\ and\ \citenamefont
  {Tanaka}}]{royall2008direct}%
  \BibitemOpen
  \bibfield  {author} {\bibinfo {author} {\bibfnamefont {C.~P.}\ \bibnamefont
  {Royall}}, \bibinfo {author} {\bibfnamefont {S.~R.}\ \bibnamefont
  {Williams}}, \bibinfo {author} {\bibfnamefont {T.}~\bibnamefont {Ohtsuka}},\
  and\ \bibinfo {author} {\bibfnamefont {H.}~\bibnamefont {Tanaka}},\
  }\bibfield  {title} {\bibinfo {title} {Direct observation of a local
  structural mechanism for dynamic arrest},\ }\href@noop {} {\bibfield
  {journal} {\bibinfo  {journal} {Nature materials}\ }\textbf {\bibinfo
  {volume} {7}},\ \bibinfo {pages} {556} (\bibinfo {year} {2008})}\BibitemShut
  {NoStop}%
\bibitem [{\citenamefont {Candelier}\ \emph {et~al.}(2010)\citenamefont
  {Candelier}, \citenamefont {Widmer-Cooper}, \citenamefont {Kummerfeld},
  \citenamefont {Dauchot}, \citenamefont {Biroli}, \citenamefont {Harrowell},\
  and\ \citenamefont {Reichman}}]{candelier2010spatiotemporal}%
  \BibitemOpen
  \bibfield  {author} {\bibinfo {author} {\bibfnamefont {R.}~\bibnamefont
  {Candelier}}, \bibinfo {author} {\bibfnamefont {A.}~\bibnamefont
  {Widmer-Cooper}}, \bibinfo {author} {\bibfnamefont {J.~K.}\ \bibnamefont
  {Kummerfeld}}, \bibinfo {author} {\bibfnamefont {O.}~\bibnamefont {Dauchot}},
  \bibinfo {author} {\bibfnamefont {G.}~\bibnamefont {Biroli}}, \bibinfo
  {author} {\bibfnamefont {P.}~\bibnamefont {Harrowell}},\ and\ \bibinfo
  {author} {\bibfnamefont {D.~R.}\ \bibnamefont {Reichman}},\ }\bibfield
  {title} {\bibinfo {title} {Spatiotemporal hierarchy of relaxation events,
  dynamical heterogeneities, and structural reorganization in a supercooled
  liquid},\ }\href@noop {} {\bibfield  {journal} {\bibinfo  {journal} {Physical
  review letters}\ }\textbf {\bibinfo {volume} {105}},\ \bibinfo {pages}
  {135702} (\bibinfo {year} {2010})}\BibitemShut {NoStop}%
\bibitem [{\citenamefont {Tanaka}\ \emph {et~al.}(2019)\citenamefont {Tanaka},
  \citenamefont {Tong}, \citenamefont {Shi},\ and\ \citenamefont
  {Russo}}]{tanaka2019revealing}%
  \BibitemOpen
  \bibfield  {author} {\bibinfo {author} {\bibfnamefont {H.}~\bibnamefont
  {Tanaka}}, \bibinfo {author} {\bibfnamefont {H.}~\bibnamefont {Tong}},
  \bibinfo {author} {\bibfnamefont {R.}~\bibnamefont {Shi}},\ and\ \bibinfo
  {author} {\bibfnamefont {J.}~\bibnamefont {Russo}},\ }\bibfield  {title}
  {\bibinfo {title} {Revealing key structural features hidden in liquids and
  glasses},\ }\href@noop {} {\bibfield  {journal} {\bibinfo  {journal} {Nature
  Reviews Physics}\ }\textbf {\bibinfo {volume} {1}},\ \bibinfo {pages} {333}
  (\bibinfo {year} {2019})}\BibitemShut {NoStop}%
\bibitem [{\citenamefont {Cubuk}\ \emph {et~al.}(2015)\citenamefont {Cubuk},
  \citenamefont {Schoenholz}, \citenamefont {Rieser}, \citenamefont {Malone},
  \citenamefont {Rottler}, \citenamefont {Durian}, \citenamefont {Kaxiras},\
  and\ \citenamefont {Liu}}]{svm-joerg}%
  \BibitemOpen
  \bibfield  {author} {\bibinfo {author} {\bibfnamefont {E.~D.}\ \bibnamefont
  {Cubuk}}, \bibinfo {author} {\bibfnamefont {S.~S.}\ \bibnamefont
  {Schoenholz}}, \bibinfo {author} {\bibfnamefont {J.~M.}\ \bibnamefont
  {Rieser}}, \bibinfo {author} {\bibfnamefont {B.~D.}\ \bibnamefont {Malone}},
  \bibinfo {author} {\bibfnamefont {J.}~\bibnamefont {Rottler}}, \bibinfo
  {author} {\bibfnamefont {D.~J.}\ \bibnamefont {Durian}}, \bibinfo {author}
  {\bibfnamefont {E.}~\bibnamefont {Kaxiras}},\ and\ \bibinfo {author}
  {\bibfnamefont {A.~J.}\ \bibnamefont {Liu}},\ }\bibfield  {title} {\bibinfo
  {title} {Identifying structural flow defects in disordered solids using
  machine-learning methods},\ }\href@noop {} {\bibfield  {journal} {\bibinfo
  {journal} {Physical Review Letters}\ }\textbf {\bibinfo {volume} {114}},\
  \bibinfo {pages} {108001} (\bibinfo {year} {2015})}\BibitemShut {NoStop}%
\bibitem [{\citenamefont {Schoenholz}\ \emph {et~al.}(2016)\citenamefont
  {Schoenholz}, \citenamefont {Cubuk}, \citenamefont {Sussman}, \citenamefont
  {Kaxiras},\ and\ \citenamefont {Liu}}]{schoenholz2016structural}%
  \BibitemOpen
  \bibfield  {author} {\bibinfo {author} {\bibfnamefont {S.~S.}\ \bibnamefont
  {Schoenholz}}, \bibinfo {author} {\bibfnamefont {E.~D.}\ \bibnamefont
  {Cubuk}}, \bibinfo {author} {\bibfnamefont {D.~M.}\ \bibnamefont {Sussman}},
  \bibinfo {author} {\bibfnamefont {E.}~\bibnamefont {Kaxiras}},\ and\ \bibinfo
  {author} {\bibfnamefont {A.~J.}\ \bibnamefont {Liu}},\ }\bibfield  {title}
  {\bibinfo {title} {A structural approach to relaxation in glassy liquids},\
  }\href@noop {} {\bibfield  {journal} {\bibinfo  {journal} {Nature Physics}\
  }\textbf {\bibinfo {volume} {12}},\ \bibinfo {pages} {469} (\bibinfo {year}
  {2016})}\BibitemShut {NoStop}%
\bibitem [{\citenamefont {Cubuk}\ \emph {et~al.}(2016)\citenamefont {Cubuk},
  \citenamefont {Schoenholz}, \citenamefont {Kaxiras},\ and\ \citenamefont
  {Liu}}]{cubuk2016structural}%
  \BibitemOpen
  \bibfield  {author} {\bibinfo {author} {\bibfnamefont {E.~D.}\ \bibnamefont
  {Cubuk}}, \bibinfo {author} {\bibfnamefont {S.~S.}\ \bibnamefont
  {Schoenholz}}, \bibinfo {author} {\bibfnamefont {E.}~\bibnamefont
  {Kaxiras}},\ and\ \bibinfo {author} {\bibfnamefont {A.~J.}\ \bibnamefont
  {Liu}},\ }\bibfield  {title} {\bibinfo {title} {Structural properties of
  defects in glassy liquids},\ }\href@noop {} {\bibfield  {journal} {\bibinfo
  {journal} {The Journal of Physical Chemistry B}\ }\textbf {\bibinfo {volume}
  {120}},\ \bibinfo {pages} {6139} (\bibinfo {year} {2016})}\BibitemShut
  {NoStop}%
\bibitem [{\citenamefont {Schoenholz}\ \emph {et~al.}(2017)\citenamefont
  {Schoenholz}, \citenamefont {Cubuk}, \citenamefont {Kaxiras},\ and\
  \citenamefont {Liu}}]{schoenholz2017relationship}%
  \BibitemOpen
  \bibfield  {author} {\bibinfo {author} {\bibfnamefont {S.~S.}\ \bibnamefont
  {Schoenholz}}, \bibinfo {author} {\bibfnamefont {E.~D.}\ \bibnamefont
  {Cubuk}}, \bibinfo {author} {\bibfnamefont {E.}~\bibnamefont {Kaxiras}},\
  and\ \bibinfo {author} {\bibfnamefont {A.~J.}\ \bibnamefont {Liu}},\
  }\bibfield  {title} {\bibinfo {title} {Relationship between local structure
  and relaxation in out-of-equilibrium glassy systems},\ }\href@noop {}
  {\bibfield  {journal} {\bibinfo  {journal} {Proceedings of the National
  Academy of Sciences}\ }\textbf {\bibinfo {volume} {114}},\ \bibinfo {pages}
  {263} (\bibinfo {year} {2017})}\BibitemShut {NoStop}%
\bibitem [{\citenamefont {Schoenholz}(2018)}]{schoenholz2018combining}%
  \BibitemOpen
  \bibfield  {author} {\bibinfo {author} {\bibfnamefont {S.~S.}\ \bibnamefont
  {Schoenholz}},\ }\bibfield  {title} {\bibinfo {title} {Combining machine
  learning and physics to understand glassy systems},\ }in\ \href@noop {}
  {\emph {\bibinfo {booktitle} {Journal of Physics: Conference Series}}},\
  Vol.\ \bibinfo {volume} {1036}\ (\bibinfo {organization} {IOP Publishing},\
  \bibinfo {year} {2018})\ p.\ \bibinfo {pages} {012021}\BibitemShut {NoStop}%
\bibitem [{\citenamefont {Wang}\ \emph {et~al.}(2020)\citenamefont {Wang},
  \citenamefont {Ding}, \citenamefont {Zhang}, \citenamefont {Podryabinkin},
  \citenamefont {Shapeev},\ and\ \citenamefont {Ma}}]{wang2020predicting}%
  \BibitemOpen
  \bibfield  {author} {\bibinfo {author} {\bibfnamefont {Q.}~\bibnamefont
  {Wang}}, \bibinfo {author} {\bibfnamefont {J.}~\bibnamefont {Ding}}, \bibinfo
  {author} {\bibfnamefont {L.}~\bibnamefont {Zhang}}, \bibinfo {author}
  {\bibfnamefont {E.}~\bibnamefont {Podryabinkin}}, \bibinfo {author}
  {\bibfnamefont {A.}~\bibnamefont {Shapeev}},\ and\ \bibinfo {author}
  {\bibfnamefont {E.}~\bibnamefont {Ma}},\ }\bibfield  {title} {\bibinfo
  {title} {Predicting the propensity for thermally activated $\beta$ events in
  metallic glasses via interpretable machine learning},\ }\href@noop {}
  {\bibfield  {journal} {\bibinfo  {journal} {npj Computational Materials}\
  }\textbf {\bibinfo {volume} {6}},\ \bibinfo {pages} {1} (\bibinfo {year}
  {2020})}\BibitemShut {NoStop}%
\bibitem [{\citenamefont {Bapst}\ \emph {et~al.}(2020)\citenamefont {Bapst},
  \citenamefont {Keck}, \citenamefont {Grabska-Barwi{\'n}ska}, \citenamefont
  {Donner}, \citenamefont {Cubuk}, \citenamefont {Schoenholz}, \citenamefont
  {Obika}, \citenamefont {Nelson}, \citenamefont {Back}, \citenamefont
  {Hassabis} \emph {et~al.}}]{bapst2020unveiling}%
  \BibitemOpen
  \bibfield  {author} {\bibinfo {author} {\bibfnamefont {V.}~\bibnamefont
  {Bapst}}, \bibinfo {author} {\bibfnamefont {T.}~\bibnamefont {Keck}},
  \bibinfo {author} {\bibfnamefont {A.}~\bibnamefont {Grabska-Barwi{\'n}ska}},
  \bibinfo {author} {\bibfnamefont {C.}~\bibnamefont {Donner}}, \bibinfo
  {author} {\bibfnamefont {E.~D.}\ \bibnamefont {Cubuk}}, \bibinfo {author}
  {\bibfnamefont {S.~S.}\ \bibnamefont {Schoenholz}}, \bibinfo {author}
  {\bibfnamefont {A.}~\bibnamefont {Obika}}, \bibinfo {author} {\bibfnamefont
  {A.~W.}\ \bibnamefont {Nelson}}, \bibinfo {author} {\bibfnamefont
  {T.}~\bibnamefont {Back}}, \bibinfo {author} {\bibfnamefont {D.}~\bibnamefont
  {Hassabis}}, \emph {et~al.},\ }\bibfield  {title} {\bibinfo {title}
  {Unveiling the predictive power of static structure in glassy systems},\
  }\href@noop {} {\bibfield  {journal} {\bibinfo  {journal} {Nature Physics}\
  }\textbf {\bibinfo {volume} {16}},\ \bibinfo {pages} {448} (\bibinfo {year}
  {2020})}\BibitemShut {NoStop}%
\bibitem [{\citenamefont {Yang}\ \emph {et~al.}(2021)\citenamefont {Yang},
  \citenamefont {Wei}, \citenamefont {Zaccone},\ and\ \citenamefont
  {Wang}}]{yang2021machine}%
  \BibitemOpen
  \bibfield  {author} {\bibinfo {author} {\bibfnamefont {Z.-Y.}\ \bibnamefont
  {Yang}}, \bibinfo {author} {\bibfnamefont {D.}~\bibnamefont {Wei}}, \bibinfo
  {author} {\bibfnamefont {A.}~\bibnamefont {Zaccone}},\ and\ \bibinfo {author}
  {\bibfnamefont {Y.-J.}\ \bibnamefont {Wang}},\ }\bibfield  {title} {\bibinfo
  {title} {Machine-learning integrated glassy defect from an intricate
  configurational-thermodynamic-dynamic space},\ }\href@noop {} {\bibfield
  {journal} {\bibinfo  {journal} {Physical Review B}\ }\textbf {\bibinfo
  {volume} {104}},\ \bibinfo {pages} {064108} (\bibinfo {year}
  {2021})}\BibitemShut {NoStop}%
\bibitem [{\citenamefont {Boattini}\ \emph {et~al.}(2020)\citenamefont
  {Boattini}, \citenamefont {Mar{\'\i}n-Aguilar}, \citenamefont {Mitra},
  \citenamefont {Foffi}, \citenamefont {Smallenburg},\ and\ \citenamefont
  {Filion}}]{boattini2020autonomously}%
  \BibitemOpen
  \bibfield  {author} {\bibinfo {author} {\bibfnamefont {E.}~\bibnamefont
  {Boattini}}, \bibinfo {author} {\bibfnamefont {S.}~\bibnamefont
  {Mar{\'\i}n-Aguilar}}, \bibinfo {author} {\bibfnamefont {S.}~\bibnamefont
  {Mitra}}, \bibinfo {author} {\bibfnamefont {G.}~\bibnamefont {Foffi}},
  \bibinfo {author} {\bibfnamefont {F.}~\bibnamefont {Smallenburg}},\ and\
  \bibinfo {author} {\bibfnamefont {L.}~\bibnamefont {Filion}},\ }\bibfield
  {title} {\bibinfo {title} {Autonomously revealing hidden local structures in
  supercooled liquids},\ }\href@noop {} {\bibfield  {journal} {\bibinfo
  {journal} {Nature communications}\ }\textbf {\bibinfo {volume} {11}},\
  \bibinfo {pages} {5479} (\bibinfo {year} {2020})}\BibitemShut {NoStop}%
\bibitem [{\citenamefont {Paret}\ \emph {et~al.}(2020)\citenamefont {Paret},
  \citenamefont {Jack},\ and\ \citenamefont {Coslovich}}]{paret2020assessing}%
  \BibitemOpen
  \bibfield  {author} {\bibinfo {author} {\bibfnamefont {J.}~\bibnamefont
  {Paret}}, \bibinfo {author} {\bibfnamefont {R.~L.}\ \bibnamefont {Jack}},\
  and\ \bibinfo {author} {\bibfnamefont {D.}~\bibnamefont {Coslovich}},\
  }\bibfield  {title} {\bibinfo {title} {Assessing the structural heterogeneity
  of supercooled liquids through community inference},\ }\href@noop {}
  {\bibfield  {journal} {\bibinfo  {journal} {The Journal of chemical physics}\
  }\textbf {\bibinfo {volume} {152}},\ \bibinfo {pages} {144502} (\bibinfo
  {year} {2020})}\BibitemShut {NoStop}%
\bibitem [{\citenamefont {Van~Kampen}(1992)}]{van1992stochastic}%
  \BibitemOpen
  \bibfield  {author} {\bibinfo {author} {\bibfnamefont {N.~G.}\ \bibnamefont
  {Van~Kampen}},\ }\href@noop {} {\emph {\bibinfo {title} {Stochastic processes
  in physics and chemistry}}},\ Vol.~\bibinfo {volume} {1}\ (\bibinfo
  {publisher} {Elsevier},\ \bibinfo {year} {1992})\BibitemShut {NoStop}%
\bibitem [{\citenamefont {Swope}\ \emph {et~al.}(2004)\citenamefont {Swope},
  \citenamefont {Pitera},\ and\ \citenamefont {Suits}}]{swope2004describing}%
  \BibitemOpen
  \bibfield  {author} {\bibinfo {author} {\bibfnamefont {W.~C.}\ \bibnamefont
  {Swope}}, \bibinfo {author} {\bibfnamefont {J.~W.}\ \bibnamefont {Pitera}},\
  and\ \bibinfo {author} {\bibfnamefont {F.}~\bibnamefont {Suits}},\ }\bibfield
   {title} {\bibinfo {title} {Describing protein folding kinetics by molecular
  dynamics simulations. 1. theory},\ }\href@noop {} {\bibfield  {journal}
  {\bibinfo  {journal} {The Journal of Physical Chemistry B}\ }\textbf
  {\bibinfo {volume} {108}},\ \bibinfo {pages} {6571} (\bibinfo {year}
  {2004})}\BibitemShut {NoStop}%
\bibitem [{\citenamefont {Prinz}\ \emph {et~al.}(2011)\citenamefont {Prinz},
  \citenamefont {Wu}, \citenamefont {Sarich}, \citenamefont {Keller},
  \citenamefont {Senne}, \citenamefont {Held}, \citenamefont {Chodera},
  \citenamefont {Sch{\"u}tte},\ and\ \citenamefont
  {No{\'e}}}]{prinz2011markov}%
  \BibitemOpen
  \bibfield  {author} {\bibinfo {author} {\bibfnamefont {J.-H.}\ \bibnamefont
  {Prinz}}, \bibinfo {author} {\bibfnamefont {H.}~\bibnamefont {Wu}}, \bibinfo
  {author} {\bibfnamefont {M.}~\bibnamefont {Sarich}}, \bibinfo {author}
  {\bibfnamefont {B.}~\bibnamefont {Keller}}, \bibinfo {author} {\bibfnamefont
  {M.}~\bibnamefont {Senne}}, \bibinfo {author} {\bibfnamefont
  {M.}~\bibnamefont {Held}}, \bibinfo {author} {\bibfnamefont {J.~D.}\
  \bibnamefont {Chodera}}, \bibinfo {author} {\bibfnamefont {C.}~\bibnamefont
  {Sch{\"u}tte}},\ and\ \bibinfo {author} {\bibfnamefont {F.}~\bibnamefont
  {No{\'e}}},\ }\bibfield  {title} {\bibinfo {title} {Markov models of
  molecular kinetics: Generation and validation},\ }\href@noop {} {\bibfield
  {journal} {\bibinfo  {journal} {The Journal of chemical physics}\ }\textbf
  {\bibinfo {volume} {134}},\ \bibinfo {pages} {174105} (\bibinfo {year}
  {2011})}\BibitemShut {NoStop}%
\bibitem [{\citenamefont {Husic}\ and\ \citenamefont
  {Pande}(2018)}]{husic2018markov}%
  \BibitemOpen
  \bibfield  {author} {\bibinfo {author} {\bibfnamefont {B.~E.}\ \bibnamefont
  {Husic}}\ and\ \bibinfo {author} {\bibfnamefont {V.~S.}\ \bibnamefont
  {Pande}},\ }\bibfield  {title} {\bibinfo {title} {Markov state models: From
  an art to a science},\ }\href@noop {} {\bibfield  {journal} {\bibinfo
  {journal} {Journal of the American Chemical Society}\ }\textbf {\bibinfo
  {volume} {140}},\ \bibinfo {pages} {2386} (\bibinfo {year}
  {2018})}\BibitemShut {NoStop}%
\bibitem [{\citenamefont {Voelz}\ \emph {et~al.}(2010)\citenamefont {Voelz},
  \citenamefont {Bowman}, \citenamefont {Beauchamp},\ and\ \citenamefont
  {Pande}}]{voelz2010molecular}%
  \BibitemOpen
  \bibfield  {author} {\bibinfo {author} {\bibfnamefont {V.~A.}\ \bibnamefont
  {Voelz}}, \bibinfo {author} {\bibfnamefont {G.~R.}\ \bibnamefont {Bowman}},
  \bibinfo {author} {\bibfnamefont {K.}~\bibnamefont {Beauchamp}},\ and\
  \bibinfo {author} {\bibfnamefont {V.~S.}\ \bibnamefont {Pande}},\ }\bibfield
  {title} {\bibinfo {title} {Molecular simulation of ab initio protein folding
  for a millisecond folder ntl9 (1- 39)},\ }\href@noop {} {\bibfield  {journal}
  {\bibinfo  {journal} {Journal of the American Chemical Society}\ }\textbf
  {\bibinfo {volume} {132}},\ \bibinfo {pages} {1526} (\bibinfo {year}
  {2010})}\BibitemShut {NoStop}%
\bibitem [{\citenamefont {No{\'e}}\ and\ \citenamefont
  {Nuske}(2013)}]{noe2013variational}%
  \BibitemOpen
  \bibfield  {author} {\bibinfo {author} {\bibfnamefont {F.}~\bibnamefont
  {No{\'e}}}\ and\ \bibinfo {author} {\bibfnamefont {F.}~\bibnamefont
  {Nuske}},\ }\bibfield  {title} {\bibinfo {title} {A variational approach to
  modeling slow processes in stochastic dynamical systems},\ }\href@noop {}
  {\bibfield  {journal} {\bibinfo  {journal} {Multiscale Modeling \&
  Simulation}\ }\textbf {\bibinfo {volume} {11}},\ \bibinfo {pages} {635}
  (\bibinfo {year} {2013})}\BibitemShut {NoStop}%
\bibitem [{\citenamefont {Mardt}\ \emph {et~al.}(2018)\citenamefont {Mardt},
  \citenamefont {Pasquali}, \citenamefont {Wu},\ and\ \citenamefont
  {No{\'e}}}]{mardt2018vampnets}%
  \BibitemOpen
  \bibfield  {author} {\bibinfo {author} {\bibfnamefont {A.}~\bibnamefont
  {Mardt}}, \bibinfo {author} {\bibfnamefont {L.}~\bibnamefont {Pasquali}},
  \bibinfo {author} {\bibfnamefont {H.}~\bibnamefont {Wu}},\ and\ \bibinfo
  {author} {\bibfnamefont {F.}~\bibnamefont {No{\'e}}},\ }\bibfield  {title}
  {\bibinfo {title} {Vampnets for deep learning of molecular kinetics},\
  }\href@noop {} {\bibfield  {journal} {\bibinfo  {journal} {Nature
  communications}\ }\textbf {\bibinfo {volume} {9}},\ \bibinfo {pages} {5}
  (\bibinfo {year} {2018})}\BibitemShut {NoStop}%
\bibitem [{\citenamefont {Xie}\ \emph {et~al.}(2019)\citenamefont {Xie},
  \citenamefont {France-Lanord}, \citenamefont {Wang}, \citenamefont
  {Shao-Horn},\ and\ \citenamefont {Grossman}}]{xie2019graph}%
  \BibitemOpen
  \bibfield  {author} {\bibinfo {author} {\bibfnamefont {T.}~\bibnamefont
  {Xie}}, \bibinfo {author} {\bibfnamefont {A.}~\bibnamefont {France-Lanord}},
  \bibinfo {author} {\bibfnamefont {Y.}~\bibnamefont {Wang}}, \bibinfo {author}
  {\bibfnamefont {Y.}~\bibnamefont {Shao-Horn}},\ and\ \bibinfo {author}
  {\bibfnamefont {J.~C.}\ \bibnamefont {Grossman}},\ }\bibfield  {title}
  {\bibinfo {title} {Graph dynamical networks for unsupervised learning of
  atomic scale dynamics in materials},\ }\href@noop {} {\bibfield  {journal}
  {\bibinfo  {journal} {Nature communications}\ }\textbf {\bibinfo {volume}
  {10}},\ \bibinfo {pages} {2667} (\bibinfo {year} {2019})}\BibitemShut
  {NoStop}%
\bibitem [{sup()}]{supp-mat}%
  \BibitemOpen
  \href@noop {} {\bibinfo {title} {See the supplemental material at --- for
  details which includes refs. [17,18,21,22,23]}}\BibitemShut {NoStop}%
\bibitem [{\citenamefont {Stukowski}(2009)}]{ovito}%
  \BibitemOpen
  \bibfield  {author} {\bibinfo {author} {\bibfnamefont {A.}~\bibnamefont
  {Stukowski}},\ }\bibfield  {title} {\bibinfo {title} {Visualization and
  analysis of atomistic simulation data with ovito--the open visualization
  tool},\ }\href@noop {} {\bibfield  {journal} {\bibinfo  {journal} {Modelling
  and Simulation in Materials Science and Engineering}\ }\textbf {\bibinfo
  {volume} {18}},\ \bibinfo {pages} {015012} (\bibinfo {year}
  {2009})}\BibitemShut {NoStop}%
\bibitem [{\citenamefont {Flenner}\ and\ \citenamefont
  {Szamel}(2015)}]{flenner2015fundamental}%
  \BibitemOpen
  \bibfield  {author} {\bibinfo {author} {\bibfnamefont {E.}~\bibnamefont
  {Flenner}}\ and\ \bibinfo {author} {\bibfnamefont {G.}~\bibnamefont
  {Szamel}},\ }\bibfield  {title} {\bibinfo {title} {Fundamental differences
  between glassy dynamics in two and three dimensions},\ }\href@noop {}
  {\bibfield  {journal} {\bibinfo  {journal} {Nature communications}\ }\textbf
  {\bibinfo {volume} {6}},\ \bibinfo {pages} {1} (\bibinfo {year}
  {2015})}\BibitemShut {NoStop}%
\bibitem [{\citenamefont {Cohen}\ and\ \citenamefont
  {Grest}(1979)}]{cohen1979liquid}%
  \BibitemOpen
  \bibfield  {author} {\bibinfo {author} {\bibfnamefont {M.~H.}\ \bibnamefont
  {Cohen}}\ and\ \bibinfo {author} {\bibfnamefont {G.}~\bibnamefont {Grest}},\
  }\bibfield  {title} {\bibinfo {title} {Liquid-glass transition, a free-volume
  approach},\ }\href@noop {} {\bibfield  {journal} {\bibinfo  {journal}
  {Physical Review B}\ }\textbf {\bibinfo {volume} {20}},\ \bibinfo {pages}
  {1077} (\bibinfo {year} {1979})}\BibitemShut {NoStop}%
\bibitem [{\citenamefont {Lad}\ \emph {et~al.}(2012)\citenamefont {Lad},
  \citenamefont {Jakse},\ and\ \citenamefont {Pasturel}}]{lad2012signatures}%
  \BibitemOpen
  \bibfield  {author} {\bibinfo {author} {\bibfnamefont {K.}~\bibnamefont
  {Lad}}, \bibinfo {author} {\bibfnamefont {N.}~\bibnamefont {Jakse}},\ and\
  \bibinfo {author} {\bibfnamefont {A.}~\bibnamefont {Pasturel}},\ }\bibfield
  {title} {\bibinfo {title} {Signatures of fragile-to-strong transition in a
  binary metallic glass-forming liquid},\ }\href@noop {} {\bibfield  {journal}
  {\bibinfo  {journal} {The Journal of chemical physics}\ }\textbf {\bibinfo
  {volume} {136}},\ \bibinfo {pages} {104509} (\bibinfo {year}
  {2012})}\BibitemShut {NoStop}%
\end{thebibliography}


%
\end{document}


\begin{titlepage}

\title{Exploring Glassy Dynamics With Markov State Models From Graph Dynamical Networks}
\author{Siavash Soltani}
\affiliation{Department of Materials Engineering, The University of British Columbia, Vancouver, BC, Canada V6T 1Z4}
\author{Chad W.~Sinclair}
\affiliation{Department of Materials Engineering, The University of British Columbia, Vancouver, BC, Canada V6T 1Z4}
\author{J\"org Rottler}
\affiliation{Department of Physics and Astronomy, The University of British Columbia, Vancouver, BC, Canada V6T 1Z1}
\affiliation{Stewart Blusson Quantum Matter Institute, The University of British Columbia, Vancouver, BC, Canada V6T 1Z4}

\maketitle
\vfill
\end{titlepage}
\widetext


\makeatletter
\renewcommand{\theequation}{\arabic{equation}}
\renewcommand{\thefigure}{\arabic{figure}}
\renewcommand{\bibnumfmt}[1]{[#1]}
\renewcommand{\citenumfont}[1]{#1}

\renewcommand{\figurename}{Figure}
\renewcommand{\thesection}{\arabic{section}} 
\renewcommand{\thesubsection}{\thesection.\arabic{subsection}}


\section{SUPPLEMENTARY INFORMATION}

\subsection{Glass models and molecular dynamics simulations.} The three-dimensional Kob-Andersen (KA) Lennard-Jones system with the composition 80-20 serves as model system detailed in the body of the paper. The interaction potential is $V_{\alpha\beta}= 4\epsilon_{\alpha\beta}[(\sigma_{\alpha\beta} / r)^{12}-(\sigma_{\alpha\beta} / r)^{6}]$ where $\epsilon_{BB}=0.5$, $\epsilon_{AB}=1.5$, $\sigma_{BB}=0.88$ and $\sigma_{AB}=0.8$. We ignore interactions with $r>2.5\sigma_{AA}$. The system has a total of 40,000 particles of type A and B in a simulation box with length of 32.18 particle diameters in x,y and z directions. The system has therefore a number density of 1.2. Results are reported in reduced units where $\sigma_{AA} \equiv \sigma$ and $\epsilon_{AA}$ the unit of energy. We first run 100,000 time steps at temperature T=1 followed by a quench to $T\approx0$ in 10 million time steps with a time step of 0.005 $\tau_{LJ}$ using a dissipative particle dynamics (DPD) thermostat. The NVE ensemble is used throughout the subsequent simulations. At each temperature, we use the saved configurations during the quench, run another $500 \tau_{LJ}$ to further relax the system at the desired temperature followed by another run during which data is taken and configurations are saved in order to be used in the GDyNets. For temperatures below $T=0.4$, the systems are held for \red{$50,000 \tau_{LJ}$} at $T=0.35$ before running the isothermal holding. This is to ensure that at temperatures where the system is not in equilibrium, aging effects are insignificant while we measure the Markov state time scale (see also Fig.~1(c)). For $T \geq 0.4$ we ensure that the system is in equilibrium by measuring the $\alpha$ relaxation time as a function of the age of glass.

\subsection{Markov State Models from Graph Dynamical Networks.} Markov Models describe the kinetics of the system as a series of memoryless jumps between states. The time evolution of the system is given by the transition matrix, $\boldsymbol{K}(\delta t)$: 
\begin{equation}
 \boldsymbol P(t+ \delta t) = \boldsymbol{K}(\delta t)^T\boldsymbol P(t)
\label{eqn:MSM}
\end{equation}
where $\boldsymbol P(t)$ is a vector whose elements are the probability of finding the system in state $i$ at time $t$.  If the system has a unique equilibrium distribution, the eigenvalues of the transition matrix have the form $|\lambda_i| \leq 1$. There is exactly one eigenvalue equal to 1 and the corresponding eigenvector represents the equilibrium distribution since the application of the transition matrix leaves this vector unchanged. Moreover, the eigenvalues of the transition matrix are real and eigenvectors form a complete set. The result is that the temporal evolution of the system in state space can be described by $i$ dynamic modes, each decaying exponentially with a characteristic relaxation time as described by equation 1 in the main text.

If the kinetics are Markovian, $\tau_{MSM}(\delta t)$ does not depend on the choice of $\delta t$. Therefore, to test the convergence to Markovianity, one should approximate $\tau_{MSM}$ as a function of lag time $\delta t$.  For the situation studied here where our states are comprised of many lumped states, achieving Markovianity will require some finite $\delta t$ so testing the evolution of $\tau_{MSM}$ as a function of $\delta t$ is important \cite{swope2004describing}.

In order to construct an efficient Markov state model from a molecular dynamics trajectory, it is necessary to reduce the system dimensionality, grouping together atoms with similar structural and dynamical features into `states'.  While in some cases the identification of these `states' is trivial, in cases like the one studied here it is not. Thus, the first step required is to identify so-called feature map functions, $\boldsymbol{\chi\left(x_t\right)}$ where $\boldsymbol x_t$ represents the local configurations of atoms at time $t$. These feature map functions then produce the vectors $\boldsymbol P(t)$ and $\boldsymbol P(t+\delta_t)$ used in Eq. (\ref{eqn:MSM}) from input time series snapshots of molecular dynamics trajectories.

To facilitate the training of the feature map functions, we have employed the GDyNets methodology originally proposed in ref. \cite{xie2019graph}. Here, we give only a short explanation focusing on details of our implementation. The GDyNet constructs graphs of input configurations for MD snapshot it is provided. Each particle is represented by a node $q_i$, which is initialized randomly according to the particle type and is a vector with a pre-defined length (64 here). Note that the local chemical environment of particles is also taken into account in feature vectors. The edges of the graph are determined by using the radial structure functions of 20 nearest neighbors of the target particle $i$,

\begin{equation}\label{eq:edges}
u_{(i,j)} = \exp({ -(d_{(i,j)}-\mu_n)^2/\gamma^2 }),
\end{equation} 
where $d_{(i,j)}$ denotes the distance between particle $i$ and its neighbor $j$,$\gamma=0.2$, $\mu_n=0.2n$ for $n=0,1,2, \cdots,35$. Therefore for each neighbor, $\mu_n$ changes from $0$ to $7$ resulting in a vector $u_{(i,j)}$ with length $36$.

The same hyperparameters as in ref. \cite{xie2019graph} are used. For each node, the vectors $q_i^{(l-1)}$, $q_j^{(l-1)}$ and $u_{(i,j)}$ from the last iteration are then concatenated:
\begin{equation}\label{eq:concat}
z_{(i,j)}^{(l-1)}= q_i^{(l-1)} \oplus q_j^{(l-1)} \oplus u_{(i,j)}.
\end{equation}
A scalar, $\alpha_{ij}$ is also calculated for each neighbor:
\begin{equation}\label{eq:alpha}
\alpha_{ij}= \frac{\exp(z_{(i,j)}^{(l-1)}W_a^{(l-1)}+b_a^{(l-1)})  }{\sum_j \exp(z_{(i,j)}^{(l-1)}W_a^{(l-1)}+b_a^{(l-1)})},
\end{equation}
which is called the attention coefficient of each neighbor.
The embedding of node $i$ is then updated using convolution:
\begin{equation}\label{eq:embdedding}
q_i^{(l)}= q_i^{(l-1)}+ \sum_j \alpha_{ij}\cdot g(z_{(i,j)}^{(l-1)}W_n^{(l-1)}+b_n^{(l-1)} )
\end{equation}
In this work, we use $l=3$ convolutional layers. $W_n$ and $b_n$ are the weight and biases of the layer and $g$ is the ReLU activation function. Note that without the attention layer, all neighbors share the same weight and bias matrices, $W_n$ and $b_n$, which neglects the differences of interaction strength between neighbors \cite{xie2019graph}. By convolution, information from particles beyond the first neighbor shell also reach the target particle. The final output of this convolutional network, $q_i^{l}$ at times $t$ and $t+\delta t$ are therefore the embeddings of atom $i$ at times $t$ and $t+\delta t$.  These results are passed to a two-layer fully-connected neural network with a Softmax activation layer in order to compute the final probability of each atom to belong to each state. Note that this two-layer network is computed using the basic framework described in \cite{mardt2018vampnets}. In this study, we use a two-state model, which is the simplest physical model one can build. Therefore the output probabilities show the probability of each atom to belong to state 0 or state 1. 
While it would be attractive to use minimize of the least square error, $|| \boldsymbol{\chi} (\boldsymbol x_{t+\delta t}) - \boldsymbol{K}^T \boldsymbol{\chi}(\boldsymbol x_t)||^2 $ as the metric against which the network can be trained, it has been shown that this is not robust \cite{mardt2018vampnets}. Rather, No\'e and co-workers have proposed the concept of the variational approach for Markov processes (VAMP) \cite{noe2013variational,mardt2018vampnets}. Here, the network has been trained to maximize the so-called VAMP-2 score based on covariance matrices obtained from the feature mapping provided by the above network at times  $t$ and ${t+\delta t}$ \cite{mardt2018vampnets}. In this way the entire trajectory can be used to optimize the network simultaneously.  

To study the dynamics in a wide range of temperatures, we have trained our network using the simulation trajectories of particles of type A (the majority particles) and type B at $T=0.4$. This trained model was then applied to simulation data obtained at other temperatures. This ensures that the transformation function $\boldsymbol{\chi}$ are identical for all temperatures. Training at different temperatures does not affect the results, see section 1.4. We chose $T=0.4$ to train the model to test the ability of the method to efficiently identify dynamic processes at temperatures where kinetics are slow. The hyperparameter $\delta t$ is chosen to be $200\tau_{LJ}$. Choosing a very small value for this hyperparameter results in overfitting (the model not being able to capture long time dynamics). A very large value is also not desirable since this causes the dynamic process to relax before we can measure its timescale. To analyze the trajectories using VAMP/GDyNets, construct the MSMs and extract $\tau_{MSM}$, the codes from \cite{gdynet} were used. 
\subsection{Generation and validation of Markov state models.}

\begin{figure}[H]
  \centering
  \subfloat{\includegraphics[width=.8\textwidth]{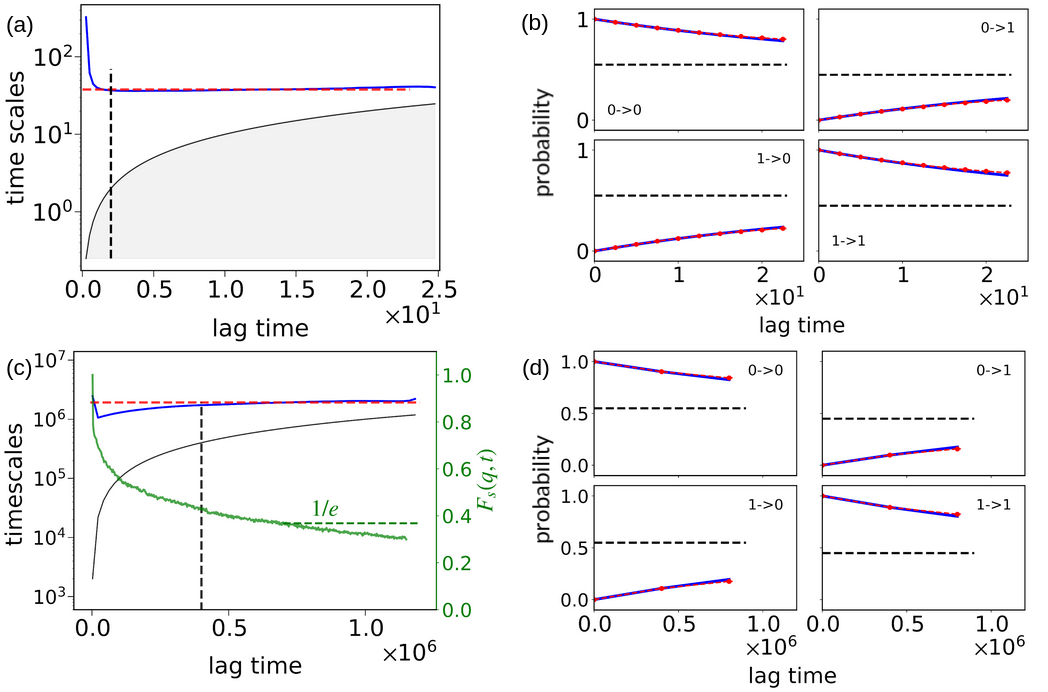}}

  \caption{Validation of Markov state models. Relaxation time as a function of lag time at (a) $T=1$ and (c) $T=0.35$. Green curve shows the self-intermediate scattering function. The time scales are obtained by averaging over 10 sets of particles by randomly dividing them in the test data. \textcolor{red}{Red dashed lines show the obtained relaxation time scales. Black dashed lines in panels (a) and (c) show the lag time $\delta t$ where the convergence to a constant timescale is observed. This lag time is used to preform CK tests, shown in b) at $T=1$ and d) at $T=0.35$. Red and blue data points in b) and d) show the left and right hand side of Eq. 6, respectively. Dashed lines show the equilibrium distribution.}}
  \label{fig:ck}
\end{figure}

In Fig.~1 we show the steps used to extract the Markov state time scale and validate the model from the trajectories. 
To test the accuracy of any Markov State model, one can estimate a transition matrix at lag times $n\delta t$, and perform a Chapman-Kolomogorov (CK) test \cite{mardt2018vampnets},
\begin{equation}\label{eq:CK}
\boldsymbol{K}(n\delta t)= \boldsymbol{K}^n(\delta t).
\end{equation}
This must hold within statistical uncertainty. For all cases studied here this test was performed, see Fig~1(b) and (d). For more information on generation and validation of Markov state models the reader is referred to \cite{prinz2011markov}.

Since the states of particles are known after the model has been trained and applied to the trajectory, we can construct the transition matrix as a function of lag time and estimate the relaxation time using the eigenvalues, see Equation (1) in the main text. Figures 1(a) and (c) show the timescale of the Markov state model as a function of lag time. The time scale becomes independent of lag time at large enough lag times and converges to a constant value (red dotted lines), indicating that the model is now Markovian, i.e. the relaxation time does not depend on lag time. We then use this lag time (black dotted lines in (a) and (c)) to construct the transition matrix and validate the Markov model with the CK test, see Equation \ref{eq:CK}. Figures (b) and (d) show this test, where the blue curves are the components of $\boldsymbol{K}^n(\delta{t})$ constructed at the convergence lag time and the red data points are the components of $\boldsymbol{K}(n\delta t)$. Note that for temperatures where the system is not in equilibrium, the Markov time scale is obtained during a time period where aging effects are minimal, i.e. the system has aged much less than one $\alpha$-relaxation time (see self-intermediate scattering function in Fig. 1(c).

\subsection{Effect of training temperature.}
In this section, we discuss the effect of the training temperature on the results. In addition to the model used in the main text, which was trained at $T=0.4$, we train the network using the trajectories at $T=0.6$ and $T=1$. Fig.~2 shows the results obtained using the three trained models. Although the three models show different values of standard deviations in the state probability distributions, all three models show similar normalized state distributions (Fig.~2(a)), Markov state time scales and correlation lengths. This behaviour of Markov state models has been observed before, where the time scale obtained using different state space discretizations converges to the same value at long times \cite{prinz2011markov}. An interesting observation is that although the length of the trajectory at $T=1$ is much shorter than that of $T=0.4$, both models give similar results. One can therefore train the network at high temperatures, where the dynamics are fast, using a short trajectory and apply this model to trajectories at lower temperatures to extract kinetic information.

\begin{figure}[t]
  \centering
  \subfloat{\includegraphics[width=1\textwidth]{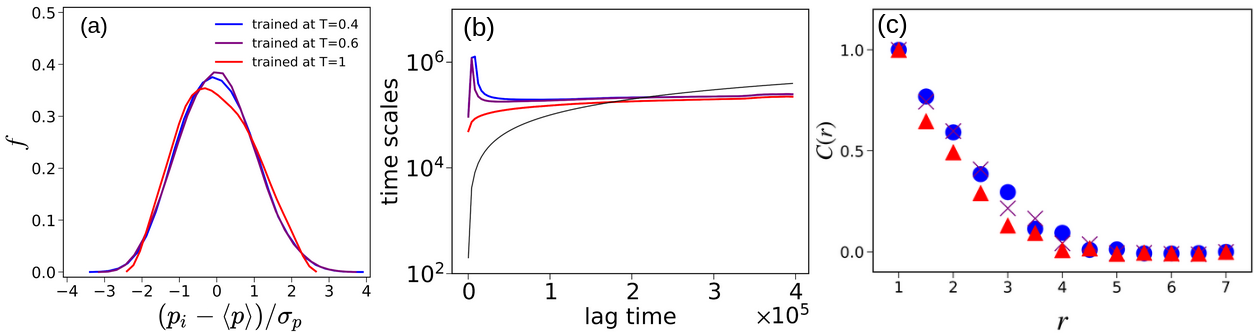}}

  \caption{Comparison between models trained using the trajectories at three different temperatures. (a) normalized probability distributions. The mean value of the distributions are $0.47$, $0.51$ and $0.52$ for models trained at $T=0.4$, $T=0.6$ and $T=1$ respectively. The standard deviations are $0.11$, $0.12$ and $0.18$. (b) Markov state time scale at $T=0.4$. (c) Correlation length of the state map at $T=0.4$.  }
  \label{fig:training}
\end{figure}

\subsection{Three-state model}

\textcolor{red}{
In this section, we show the results obtained by training a three-state model on the trajectory at $T=0.4$. Fig. 3 shows an example of a state map with particles colored according to their probability of being in each of the three states. State 0 and state 2 (Fig. 3(a) and (c)) reveal the same features as the two-state model but exhibit a somewhat larger contrast. The new intermediate state is mostly featureless: all particles have the same probability 0.5 except those whose probability is close to one in either state 0 or 2. For those particles, the probability of being in state 1 must be depressed by conservation of probability. There are now two dynamic modes, and their relaxation time scales are shown in Fig.~3(d). The greater relaxation time scale is equal to the time scale obtained already by a two-state model (also shown in Fig.~2(d)). This observation is agreement with the theory of Markov state models \cite{prinz2011markov}, from which we learn that adding more states leads to finding subdominant transitions with smaller time scales. The CK tests of Fig.~3(e) confirm that the model is meaningful and well converged. Apart from the equilibrium distribution, two eigenvectors can be found, which are shown in Fig.~3(f). These eigenvectors provide insight into the processes associated with the different timescales. We see, for instance, that the slowest mode corresponds to transitions between states 0 and 1 into state 2, while the second faster timescale corresponds to transitions between states 0 and 2 in and out of the intermediate state 1.    }

\begin{figure}[H]
  \centering
  \subfloat{\includegraphics[width=1\textwidth]{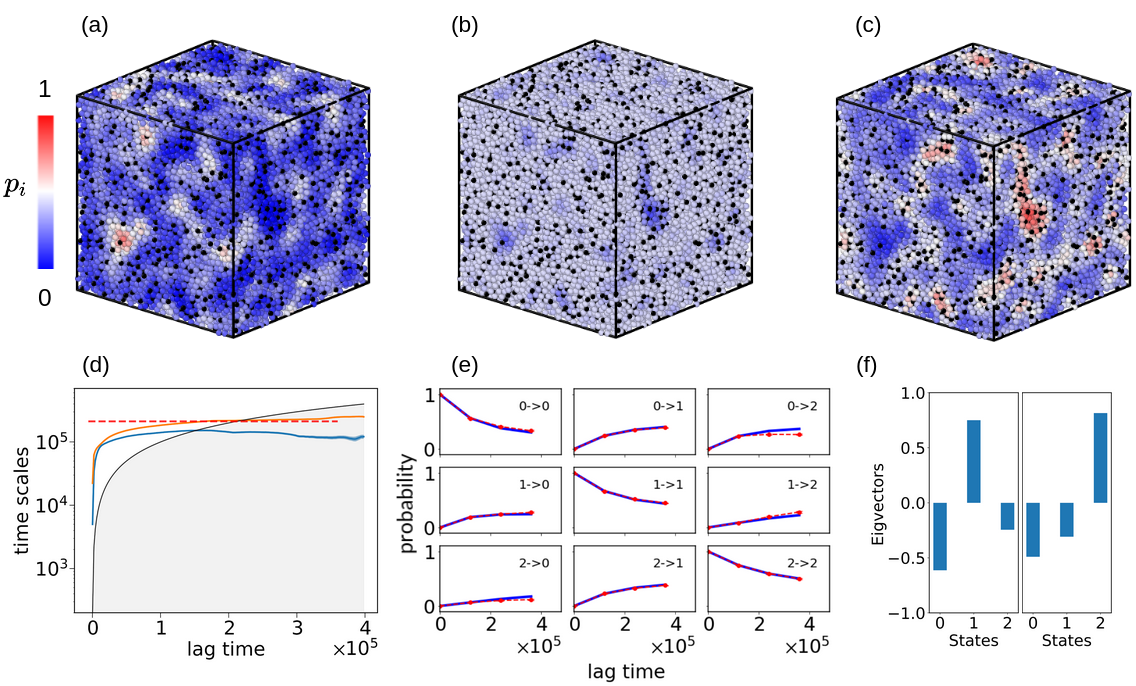}}

  \caption{ Three state model at $T=0.4$. Particles $i$ are colored according to their probability $p_i$ of being in a) state $0$, b) state $1$ and c) state $2$. d) Relaxation timescales of dynamic modes. Red dashed line shows the timescale obtained from the two-state model. e) CK tests as in Fig. 1 and d) Second and third eigenvectors.   }
  \label{fig:3states}
\end{figure}